# Quantum Anomalous Hall and Half-metallic Phases in Ferromagnetic (111) Bilayers of 4d and 5d Transition Metal Perovskites

Hirak Kumar Chandra[1,2] and Guang-Yu Guo[1,3]

[1] *Department of Physics, National Taiwan University, Taipei 10617, Taiwan*
[2] *Department of Physics, Heritage Institute of Technology, Kolkata 700107, India*
[3] *Physics Division, National Center for Theoretical Sciences, Hsinchu 30013, Taiwan*

Extraordinary electronic phases can form in artificial oxide heterostructures, which will provide a fertile ground for new physics and also give rise to novel device functions. Based on a systematic first-principles density functional theory study of the magnetic and electronic properties of the (111) superlattices $(ABO_3)_2/(AB'O_3)_{10}$ of 4d and 5d transition metal perovskite (B = Ru, Rh, Ag, Re, Os, Ir, Au; $AB'O_3$ = $LaAlO_3$, $SrTiO_3$), we demonstrate that due to quantum confinement, bilayers $(LaBO_3)_2$ (B = Ru, Re, Os) and $(SrBO_3)_2$ (B = Rh, Os, Ir) are ferromagnetic with ordering temperatures up to room temperature. In particular, bilayer $(LaOsO_3)_2$ is an exotic spin-polarized quantum anomalous Hall insulator, while the other ferromagnetic bilayers are metallic with large Hall conductances comparable to the conductance quantum. Furthermore, bilayers $(LaRuO_3)_2$ and $(SrRhO_3)_2$ are half-metallic, while bilayer $(SrIrO_3)_2$ exhibits peculiar colossal magnetic anisotropy. Our findings thus show that 4d and 5d metal perovskite (111) bilayers are a class of quasi-two-dimensional materials for exploring exotic quantum phases and also for advanced applications such as low-power nanoelectronics and oxide spintronics.

## I. Introduction

In 1980, it was discovered [1] that when a strong perpendicular magnetic field is applied to a two-dimensional(2D) electron gas at low temperatures, the Hall conductance is quantized in units of the conductance quantum ($e^2/h$) due to Landau-level quantization. This integer quantum Hall effect (IQHE) is subsequently found to be directly connected with the topological property of the 2D bulk insulating states, characterized by a topological invariant called the Chern number [2,3]. This topological understanding of the IQHE implies that the IQHE can also occur in other time-reversal symmetry broken systems with a topologically non-trivial band structure in the absence of the external magnetic field and Landau levels, such as a ferromagnetic (FM) insulator, leading to the so-called quantum anomalous Hall effect (QAHE). This effect was first proposed by Haldane in a honeycomb lattice model with a staggered magnetic field that produces zero average flux per unit cell [4]. Due to its intriguing nontrivial topological properties and fascinating potential application for designing dissipationless electronics and spintronics, extensive theoretical studies have been made recently to search for real materials to harbor such QAHE. Indeed, specific material systems such as FM quantum wells in the insulating state [5], FM topological insulator (TI) films [6], graphene on magnetic substrates [7,8], and noncoplanar antiferromagnetic layered oxide [9] have been predicted.

Importantly, based on the prediction in Ref. [6], the QAHE has recently been observed in the Cr-doped $(Bi,Sb)_2Te_3$ films [10]. Nevertheless, the QAH phase appears at extremely low tempetures due to the small band gap, weak magnetic coupling and low carrier mobility in the sample. This hinders further exploration of the exotic properties of the Chern insulator and also its applications. The low carrier mobility could result from the disorder due to the doped magnetic impurities in the sample, while the weak magnetic coupling could stem from the localized Cr 3d orbitals which hardly overlap with the orbitals on the neighboring sites. The problems of the weak magnetic coupling and small band gap could be overcome by adopting 4d and 5d transition metal atoms which simultaneously have more extended d orbitals and stronger spin-orbit coupling (SOC). Therefore, it would be fruitful to search for high temperature QAH phase in stoichiometry ferromagnetic 4d and 5d transition metal compounds.

Transition metal oxides (TMOs) span a wide range of crystalline structures and exhibit a rich variety of fascinating properties such as charge-orbital ordering, high temperature superconductivity, colossal magnetoresistance and half-metallic behavior [11]. Artificial atomic scale TMO heterostructures offer the prospect of greatly enhancing these fascinating properties or of combining them to realize novel properties and functionalities [12,13] such as the conductive interface between two insulating oxides [14,15]. Very recently, the topology of the electronic band structure of TMO heterostructures has also been investigated. [16-19]. In particular, (111) perovskite superlattices of insulating and metallic oxides were recently predicted to be TIs, based on the tight-binding (TB) modelling and first-principles calculations. [16] Furthermore, it was also predicted that the QAH phase could occur in (001) double-perovskite $La_2MnIrO_6$ monolayer [20] and (111) double-perovskite $La_2FeMoO_6$ bilayer [18] as well as (111) Ir oxide superlattice [21] and (111) $LaCoO_3$ bilayer [22].

Here we present a systematic first-principles density functional theory study of the magnetic and electronic properties of a number of (111) bilayers [$(ABO_3)_2$] of 4d and 5$d$ transition metal (B) perovskites (B = Ru, Rh, Ag, Re, Os, Ir, Au) embedded in an insulating perovskite (AB'O$_3$) matrix of either LaAlO$_3$ (LAO) or SrTiO$_3$ (STO). We find that bilayers (LaBO$_3$)$_2$ (B = Ru, Re, Os) and (SrBO$_3$)$_2$ (B = Rh, Os, Ir) are ferromagnetic with high Curie temperatures and large magnetic anisotropy energies. Fascinatingly, bilayer (LaOsO$_3$)$_2$ is a rare spin-polarized Chern insulator with spin-polarized, dissipationless edge currents tunable by a magnetic field, thus being superior for future spintronics. The other ferromagnetic bilayers are metallic with large anomalous Hall conductances comparable to the conductance quantum, thereby being promising for magnetic sensors. Moreover, bilayers (LaRuO$_3$)$_2$ and (SrRhO$_3$)$_2$ are half-metallic with nearly 100 % spin-polarization, thus being useful for magneto-transport devices. We thus demonstrate that 4$d$ and 5$d$ metal perovskite (111) bilayers are a class of quasi-2D materials for exploring novel electronic phases and also for technological applications such as low-power nanoelectronics and oxide spintronics.

Table I: Candidate 4$d$ and 5$d$ transition metal perovskites.

|  | Config. | Bulk | superlattice |
| --- | --- | --- | --- |
| LaRuO$_3$ | $d^5$ ($t_{2g}^5$) | metallic[23] | - |
| LaAgO$_3$ | $d^8$ ($t_{2g}^2$) | metallic(cal.)[24] | - |
| LaReO$_3$ | $d^4$ ($t_{2g}^4$) | metallic(cal.)[a] | - |
| LaOsO$_3$ | $d^5$ ($t_{2g}^5$) | metallic(cal.)[a] | - |
| LaAuO$_3$ | $d^8$ ($e_g^2$) | metallic[26,27] | - |
| SrRhO$_3$ | $d^5$ ($t_{2g}^5$) | metallic[28] | (Ref. [29]) |
| SrAgO$_3$ | $d^7$ ($e_g^1$) | metallic(cal.)[a] | - |
| SrOsO$_3$ | $d^4$ ($t_{2g}^4$) | metallic[30] | - |
| SrIrO$_3$ | $d^5$ ($t_{2g}^5$) | metallic[31,32] | (Ref. [33]) |

[a] Calculations (see Fig. S1 in Ref. [25]).

## II. Structure and Methods

We consider transition metal perovskite bilayers (ABO$_3$)$_2$ sandwiched by an insulating perovskite AB'O$_3$ (AB'O) slab in the (ABO$_3$)$_2$/(AB'O)$_{10}$ superlattices grown along the [111] direction, where B is a 4$d$ or 5$d$ transition metal with an open $t_{2g}$ or $e_g$ shell (see Table I). The resultant superlattices have a trigonal symmetry, and the B atoms in each bilayer form a buckled honeycomb lattice (see Fig. 1). Note that transition metal (B) atoms in the bilayer form a buckled honeycomb lattice (see Fig. 1). Since the AB'O slab would be much thicker than the ABO$_3$ bilayer, the AB'O slab could be regarded as the substrate. Therefore, as in Ref. [16], the superlattice lattice constants are set to $a = \sqrt{2}\,a_0$ and $c = 4\sqrt{3}\,a_0$ where $a_0$ is the cubic lattice constant of the AB'O perovskite. With the lattice constants $a$ and $c$ fixed, the internal coordinates of all the atoms in each superlattice are theoretically optimized, as described next.

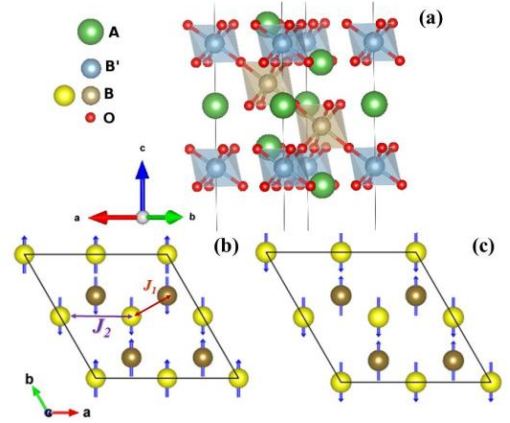

Fig. 1: Atomic and magnetic structures of a (ABO$_3$)$_2$/(AB'O$_3$)$_{10}$ superlattice. (a) Part of the unit cell containing the (ABO$_3$)$_2$ bilayer, (b) z-AF and (c) i-AF magnetic configuration in the buckled honeycomb lattice formed by the B atoms. Arrows represent the directions of spin moments on the B atoms. Two colors denote the B atoms on the two different planes in the bilayer, which form a buckled honeycomb lattice.

The cubic lattice constant $a_0$ of LaAlO$_3$ (LAO) and SrTiO$_3$ (STO) is determined theoretically by density functional theory calculations with the generalized gradient approximation to the exchange-correlation potential (GGA) [34]. The GGA calculations are carried out by using the accurate projector augmented wave (PAW) method [35], as implemented in the VASP code [36-38]. A plane wave cutoff energy of 400 eV is used throughout. The calculated lattice constant $a_0$ of LaAlO$_3$ (SrTiO$_3$) is 3.81 Å (3.94 Å), being close to the experimental value of 3.79 Å [39] (3.91 Å [40]). With the lattice constants $a$ and $c$ fixed, the internal coordinates of all the atoms in each superlattice are relaxed untill the total energy change is within $10^{-6}$ eV. In these atomic structural optimizations, a $k$-point mesh of 12×12×2 is used for the Brilloin zone integration.

Spin-polarized electronic band structure calculations for all the superlattices are performed by using the theoretical atomic structures. We consider the FM, zigzag antiferromagnetic (AF) (z-AF) [Fig. 1(b)] and interlayer AF (i-AF) [Fig. 1(c)] structures. To understand the magnetic instability and also estimate the magnetic ordering temperature ($T_C$) in the magnetic bilayers, we evaluate the first-neighbor ($J_1$) and second-neighbor ($J_2$) inter-atomic exchange coupling parameters by mapping the calculated total energies of the FM, z-AF and i-AF magnetic configurations to the classical Heisenberg model $H = E_0 - \sum_{i>j} J_{ij} \sigma_i \cdot \sigma_j$ where $J_{ij}$ is the exchange coupling parameter between sites $i$ and $j$, $\sigma_i$ denotes the direction of spin on site $i$.

Table II. Calculated magnetic and electronic properties of the FM $(ABO_3)_2$ bilayers in the $(ABO_3)_2/(AB'O)_{10}$ superlattices. $m_s^t$ and $m_s^B$ denote the total and B-atomic spin magnetic moments, respectively. $\Delta E_{ma}$, $E_g$, $J_1$ and $J_2$ represent the MAE, band gap, nearest neighbor and second neighbor exchange coupling parameters, respectively. A positive (negative) $\Delta E_{ma}$ means that the easy axis is along the $c$-axis (lies in-plane). $N(E_F)$, $P$ and $\sigma_{xy}^A$ represent the density of states at the Fermi level, its spin polarization and the anomalous Hall conductivity, respectively. A zero band gap means that the system is metallic. The values of $m_s^t$ and $P$ in brackets are obtained without SOC.

| Bilayer | $J_1/J_2$ (meV) | $T_C$ (K) | $m_s^t$ ($\mu_B$/cell) | $m_s^B$ ($\mu_B$/atom) | $\Delta E_{ma}$ (meV/Cell) | $E_g$ (meV) | $N(E_F)$ (states/eV/cell) | $P$ | $\sigma_{xy}^A$ ($e^2$/hc) |
|---|---|---|---|---|---|---|---|---|---|
| $(LaRuO_3)_2$ | 25.6/0.6 | 311 | 1.94 (2.00) | 0.70 | 2.4 | 0 | 10.41 | -0.98 (-1.00) | 1.22 |
| $(LaReO_3)_2$ | 2.4/1.2 | 56 | 0.92 (1.49) | 0.33 | -0.20 | 0 | 12.01 | -0.08 (0.56) | 1.78 |
| $(LaOsO_3)_2$ | 22.1/1.7 | 296 | 0.83 (2.00) | 0.30 | 2.5 | 38 | - | - (-1.00) | 2.00 |
| $(SrRhO_3)_2$ | 20.6/2.0 | 285 | 1.89 (2.00) | 0.54 | -1.8 | 0 | 7.98 | -0.93 (-1.00) | 1.84 |
| $(SrOsO_3)_2$ | 15.9/4.0 | 277 | 2.66 (3.88) | 0.81 | 2.1 | 0 | 17.25 | 0.18 (0.78) | 1.52 |
| $(SrIrO_3)_2$ | 18.5/2.0 | 263 | 0.16 (2.00) | 0.06 | ∞ | 0 | 1.00 | 0.27 (-1.00) | 0.26 |

The relativistic SOC, the coupling between the spin of electron and the fictitious magnetic field created by its own orbital motion around the nucleus, is the fundamental cause of many intriguing properties of magnetic solids such as anomalous Hall effect (AHE) [41], magnetocrystalline anisotropy energy (MAE) [42] and QAH insulating phase [10]. Therefore, we further perform the band structure calculations with the SOC included for all the superlattices. Again, the $k$-point mesh of ($\sigma_{xy}^A$) is used for the Brilloin zone integration and the magnetization is assumed to be along the $c$-axis unless stated otherwise. The MAE of a magnetic solid is defined as the total energy difference between two different magnetization directions. Here, the MAE is calculated by taking the magnetization along the $c$-axis and $a$-axis (the [11-2]) direction). A fine $k$-point mesh of 20×20×4 is used, which ensures that the calculated MAEs are converged within a few percents.

The anomalous Hall conductivity (AHC) for each magnetic superlattice is also calculated by using Berry phase formalism. [43] Within this formalism, AHC ($\sigma_{xy}^A$) is given by a sum of the Berry curvature over all $k$-points within the Brillouin zone for all the occupied bands. Since a large number of $k$-points are needed to get accurate AHCs, we use the efficient Wannier interpolation [44, 45] method based on maximally localized Wannier functions (MLWFs) [46]. Since the energy bands around the Fermi level are dominated by mainly B $d$-orbitals, twenty MLWFs per unit cell of B ($d^\uparrow$, $d^\downarrow$) orbitals are constructed by fitting to the GGA band structure in the energy window from -2.2 eV to 1.2 eV. The band structure obtained by the Wannier interpolation for the $(LaOsO_3)_2/(LAO)_{10}$ superlattice agrees well with that from the GGA calculation (see Fig. S2 in Ref. [25]). The AHC ($\sigma_{xy}^A$) for all the superlattices was then evaluated by taking a very dense $k$-point mesh of 144×144×12 in the Brillouin zone.

## III. Results and Discussion

*Magnetism.*—Among the considered perovskite (111) bilayers $(ABO)_2$ in the $(ABO)_2/(AB'O)_{10}$ superlattices, bilayers $(LaBO)_2$ (B = Ru, Re, Os) and $(SrBO)_2$ (B = Rh, Os, Ir) are found to be ferromagnetic. The calculated principal properties of these FM (111) bilayers are listed in Table II. All the FM bilayers except $(SrIrO)_2$, have a considerable magnetization (see Table II). Interestingly, bilayer $(LaOsO)_2$ is a FM insulator, while the other FM bilayers are metallic with a large density of states at the Fermi level ($E_F$) except $(SrIrO)_2$, which is a semimetal with a pseudo-gap and hence a small density of states at the $E_F$. We notice that all these FM bilayers have an open B $t_{2g}$ shell (see Table I). The other considered bilayers are nonmagnetic and thus are not listed in Table II. It is somewhat surprising that all the bilayers having an open $e_g$ shell (see Table I) are nonmagnetic. In particular, Ref. [16] predicted that bilayer $(LaAuO)_2$ could be a FM Chern insulator.

Table II shows that all the calculated exchange coupling parameters are positive (i.e., FM coupling) and hence the magnetic structure in these bilayers is ferromagnetic. Moreover, all the $J_1$ values are rather large because 4$d$ and 5$d$ orbitals are rather extended, as mentioned before, and thus give rise to large hybridizations between the $d$ orbitals on the neighboring magnetic atoms. Based on these exchange coupling parameters, a mean-field estimation (see, e.g., Ref. [47]) would lead to magnetic ordering temperatures ranging from 56 K to room temperature (311 K) (Table II).

The MAE of a magnetic solid determines how strong the magnetization in the solid could be pinned along its easy axis. Thus, from the application point of view, the MAE is an important property for a nanomaterial because, e.g., a large MAE would help reduce thermal (superparamagnetic) fluctuations considerably. Table II shows that many of the FM bilayers have a large MAE, reflecting the strong SOC of 4$d$ and 5$d$ transition metal atoms as well as the geometric anisotropy of the bilayers. Note that the MAE for bulk Fe and Ni is ~5.0 $\mu$eV/atom [48]. Interestingly, bilayers

(LaRuO)$_2$, (LaOsO)$_2$ and (SrOsO)$_2$ has a perpendicular anisotropy and their MAEs are comparable to that of the ordered L1$_0$ FePt alloy (~1.0 meV/FePt) [49], which has the largest MAE among the transition metal alloys. Clearly, these bilayers are promising materials for high density magnetic data storage devices. Remarkably, in bilayer (SrIrO)$_2$, the magnetization vanishes when it is rotated to an in-plane direction, i.e., it is impossible to rotate the magnetization to an in-plane direction, a peculiar phenomenon dubbed colossal magnetic anisotropy [50], and thus the MAE of this bilayer is marked as MAE = ∞ in Table II.

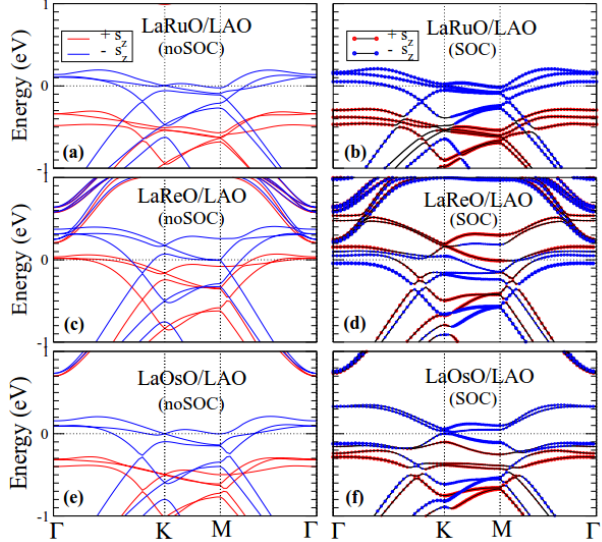

Fig. 2: Band structures of the (LaRuO)$_2$/(LAO)$_{10}$ (a-b), (LaReO)$_2$/(LAO)$_{10}$ (c-d) and (LaOsO)$_2$/(LAO)$_{10}$ (e-f) superlattices. Zero refers to the Fermi level.

*Band structure.*—Among Band structures calculated without and with the SOC for the FM (LaBO)$_2$/(LAO)$_{10}$} (B = Ru, Re, Os) superlattices are displayed in Fig. 2, and that for the FM (SrBO)$_2$/(STO)$_{10}$ (B = Rh, OS, Ir) are shown in Fig. 3. In the absence of the SOC, all the FM superlattices are metallic. Furthermore, bilayers (LaBO)$_2$ (B = Ru, Os) and (SrBO)$_2$ (B = Rh, Ir) are half-metallic, with the energy bands in the vicinity of the $E_F$ being purely spin-down (Figs. 2 and 3). This half-metallicity is consistent with the integer values of spin magnetic moments (2.0 $\mu_B$/cell) (Table II). Interestingly, there is a Dirac point made up of spin-down bands near the $E_F$ located at the K point in the (LaOsO)$_2$ bilayer, as could be expected from a honeycomb lattice [16].

The SOC should have significant effects on the electronic and magnetic properties of these 4d and 5d transition metal perovskite superlattices. Indeed, the strong SOC significantly reduces the spin magnetic moment in (SrIrO$_3$)$_2$/(STO$_3$)$_{10}$ from 2.0 to 0.16 $\mu_B$/cell (Table II). Consequently, the band spin-splittings become small [see Fig. 3(f)], and the superlattice becomes a semimetal, with tiny hole (electron) pockets at the Γ (K) point. Remarkably, a band gap is opened near the Dirac point at the K point in (LaOsO)$_2$/(LAO)$_{10}$ when the SOC is turned on [see Fig. 2(f)], although the energy bands just below and above the gap remain of spin-down character. In fact, this band gap is topologically nontrivial, as will be shown below. Nevertheless, in the presence of the SOC, the other four superlattices remain metallic with several d bands crossing The $E_F$ and hence a large density of states (see Figs. 2 and 3 as well as Table II). Furthermore, Fig. 2(b) and Fig. 3(b) show that the energy bands in the vicinity of the $E_F$ In (LaRuO)$_2$/(LAO)$_{10}$ and (SrIrO$_3$)$_2$/(STO)$_{10}$ remain almost purely spin-down, i.e., these superlattices remain half-metallic. Indeed, the calculated spin-polarization $P$ is -0.98 for bilayer (LaRuO)$_2$ and -0.93 for bilayer (SrRhO)$_2$, and the calculated spin moments of 1.94 and 1.89 $\mu_B$/cell are close to the quantized value (2.0 $\mu_B$/cell) (see Table II).

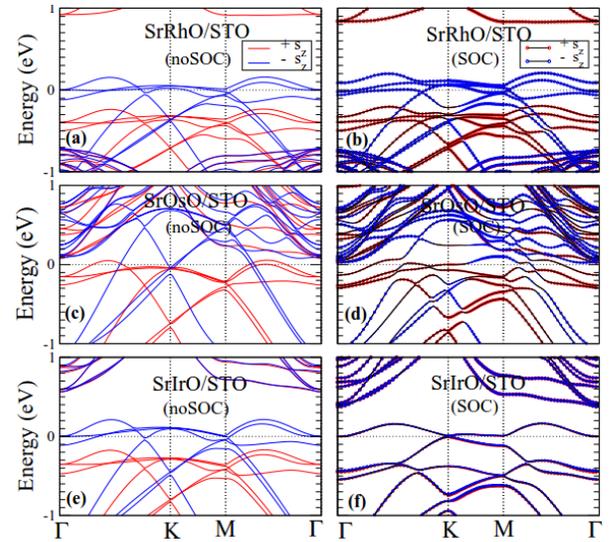

Fig. 3: Band structures of the (SrRhO)$_2$/(STO)$_{10}$ (a-b), (SrOsO)$_2$/(STO)$_{10}$ (c-d) and (SrIrO)$_2$/(STO)$_{10}$ (e-f) superlattices. Zero refers to the Fermi level.

*Quantum confinement.*—In the FM superlattices listed in Table II, each metal perovskite bilayer is sandwiched by an insulating perovskite slab. Thus, the conduction bands made up of mainly transition metal $t_{2g}$ orbitals are confined within the metal perovskite bilayer. In (LaOsO)$_2$/(LAO)$_{10}$, for example, the 12 conduction bands dominated by Os $t_{2g}$ orbitals are separated from Os $e_g$-dominated upper conduction bands by a small gap of 0.4 eV and also from O p-dominated valence bands by a gap of 0.6 eV (see Fig. 4). The width of this $t_{2g}$ band manifold is 2.4 eV and is significantly smaller than that (3.7 eV) of the Os $t_{2g}$ conduction bands in the cubic LaOsO perovskite [see Fig. S1 (b) in Ref. [25]]. The band width would be 2.2 eV from the relativistic nonmagnetic calculation. This significant narrowing of the conduction band due to the quantum confinement, leads to enhanced exchange interaction among the $t_{2g}$ electrons and results in the formation of the FM state in the superlattices listed in Table II. Figure 5 show the calculated charge density and spin density of the conduction electrons near the $E_F$ in $(LaRuO)_2/(LAO)_{10}$, which are clearly confined within the $(LaRuO)_2$ bilayer. Therefore, the electron systems in these FM superlattices are in fact quasi-2D. Furthermore, the $(LaRuO)_2/(LAO)_{10}$ and $(SrRhO)_2/(STO)_{10}$ superlattices are exotic quasi-2D half-metallic systems. We note that although many compounds have been predicted to be half-metals and the half-metallicity has been unequivocally observed only in $CrO_2$ [51], quasi-2D half-metallic systems are rare [52].

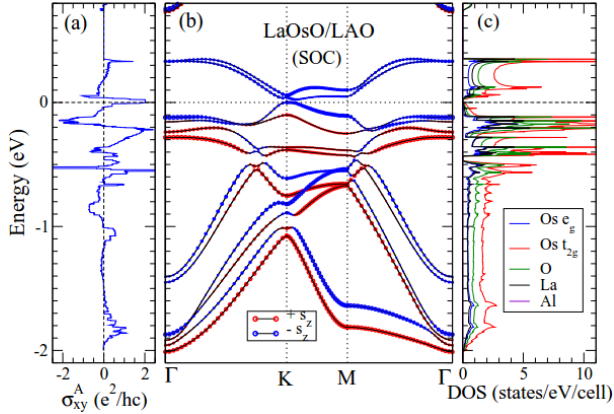

Fig. 4: (a) Anomalous Hall conductivity ($\sigma^A_{xy}$), (b) band structure and (c) atom- and Os $d$ orbital-decomposed densities of states (DOS) of the $(LaOsO)_2/(LAO)_{10}$ superlattice. Zero refers to the top of valence bands.

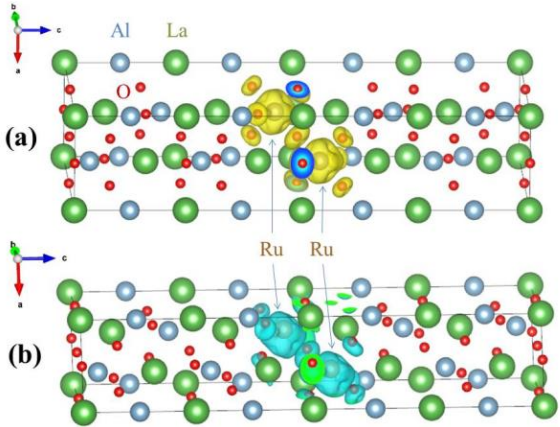

Fig. 5: (a) Charge and (b) spin density distributions of conduction electrons in the $(LaRuO)_2/(LAO)_{10}$ superlattice. Both charge and spin densities are confined within the $(LaRuO)_2$ bilayer located at the central part.

*Quantum anomalous Hall phase.*—As mentioned before, the FM $(LaOsO_3)_2/(LAO)_{10}$ superlattice is found to be a semiconductor with the insulating gap opened near the Dirac point at the K point when the SOC is included. We thus could expect that the band gap would be topologically nontrivial and the $(LaOsO_3)_2$ bilayer is a Chern insulator. To verify the topological nature of this insulating gap, we calculate the AHC ($\sigma^A_{xy}$) for this superlattice. For a three-dimensional (3D) quantum Hall insulator, $\sigma^A_{xy} = n\ e^2/hc$ where $c$ is the lattice constant along the $c$-axis normal to the plane of longitudinal and Hall currents and $n$ is an integer known as the Chern number ($n_C$) [53]. For a normal FM insulator, on the other hand, $\sigma^A_{xy} = 0$. The calculated AHC of $(LaOsO_3)_2/(LAO)_{10}$ is displayed in Figs. 4(a) and S2(b) as a function of $E_F$. Figure 4(a) indeed shows that the $\sigma^A_{xy}$ is 2.0 $e^2$/hc within the band gap, being quantized with the Chern number $n_C = 2$. We note that for $(LaOsO_3)_2/(LAO)_{10}$, the estimated Curie temperature is around room temperature ($T_C = 296$ K) and, as mentioned before, the calculated MAE is extremely large (Table II), which would help suppress thermal fluctuations significantly at high temperatures. Thus, we have the possibility of achieving a high temperature QAH phase in the FM $(LaOsO_3)_2$.

We notice that the anomalous Hall effect, discovered in 1881 by Hall [54], has recently received renewed interest [41] because it is an archetypal spin-related transport phenomenon. Moreover, FM materials with a large AHC could find such technological applications as magnetization sensors [55]. We thus calculate the AHC also for the other FM superlattices. Table II shows that bilayers $(LaReO_3)_2$, $(SrRhO_3)_2$ and $(SrOsO_3)_2$ have a large AHC, being close to that of bilayer $(LaOsO_3)_2$, and hence they are promising materials for, e.g., magnetic sensors. Furthermore, bilayer $(SrRhO_3)_2$ is half-metallic and its Hall current would be highly spin-polarized, thus having high application potential for low-power consumption spintronic devices.

To ensure that the QAH phase found in bilayer $(LaOsO_3)_2$ is robust against the variation of the thickness of the insulating LAO slab in the $(LaOsO_3)_2/(LAO)_{10}$ superlattice, we further perform the calculations for $(LaOsO_3)_2/(LAO)_7$ and $(LaOsO_3)_2/(LAO)_{13}$. We find that both superlattices have a nearly identical band structure to that of the $(LaOsO_3)_2/(LAO)_{10}$ (see Fig. S3 in Ref. [25]). In the former case, the band gap is reduced to ~10 meV and the magnetization to 0.66 $\mu_B$/cell. In the latter case, however, a larger band gap of 44 meV, together with a magnetization of 0.85 $\mu_B$/cell, is found. Finally, we note that compared to the QAH phases predicted so far in other real materials [56], the QAH phase in bilayer $(LaOsO_3)_2$ has one distinct feature. The toplogical band gap was opened in the purely spin-down bands [see Fig. 2(e) and 2(f)]. Thus, it is a spin-polarized QAH phase and the edge current would be spin-polarized [57]. Moreover, both the spin-polarization and the direction of the edge current can be reversed by reversing the magnetization via an applied magnetic field. Therefore, the QAHE predicted in bilayer $(LaOsO_3)_2$ would be superior for future spintronic applications.

**Acknowledgments**: This work is supported by the Ministry of Science and Technology, National Center for Theoretical Sciences and Academia Sinica of The R.O.C.

**Supplemental Material for "Quantum anomalous Hall and half-metallic phases in ferromagnetic (111) bilayers of *4d* and *5d* transition metal perovskites"**

I. Supplementary Figures

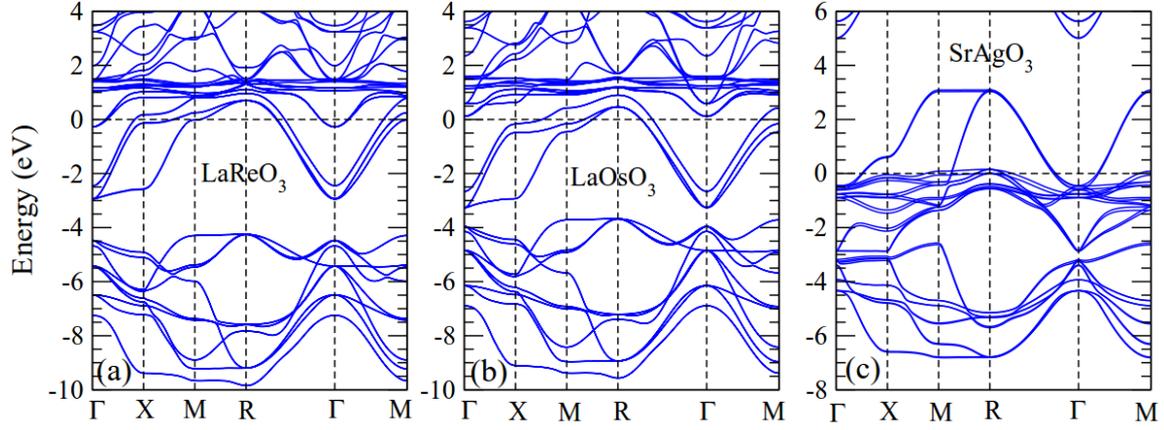

Figure S1. Relativistic band structures of cubic perovskites LaReO$_3$ (a), LaOs$_3$ (b) and SrAgO$_3$ (c) calculated using the theoretical lattice constats. Calculated density of states at the Fermi level (0 eV) is 3.66 states/eV/cell for LaReO$_3$, 2.52 states/eV/cell for LaOsO$_3$, and 2.96 states/eV/cell for SrAgO$_3$. Note that SrAgO$_3$ is a weak ferromagnet with a spin magnetic moment of 0.23 $\mu_B$/cell.

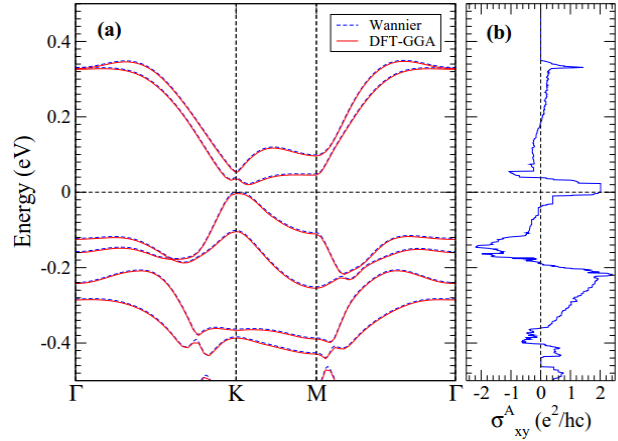

Figure S2. (a) *Ab initio* GGA and Wannier-interpolated band structures and (b) anomalous Hall conductivity ($\sigma^A_{xy}$) of the (LaOsO)$_2$/(LAO)$_{10}$ superlattice. The Fermi level is at 0 eV.

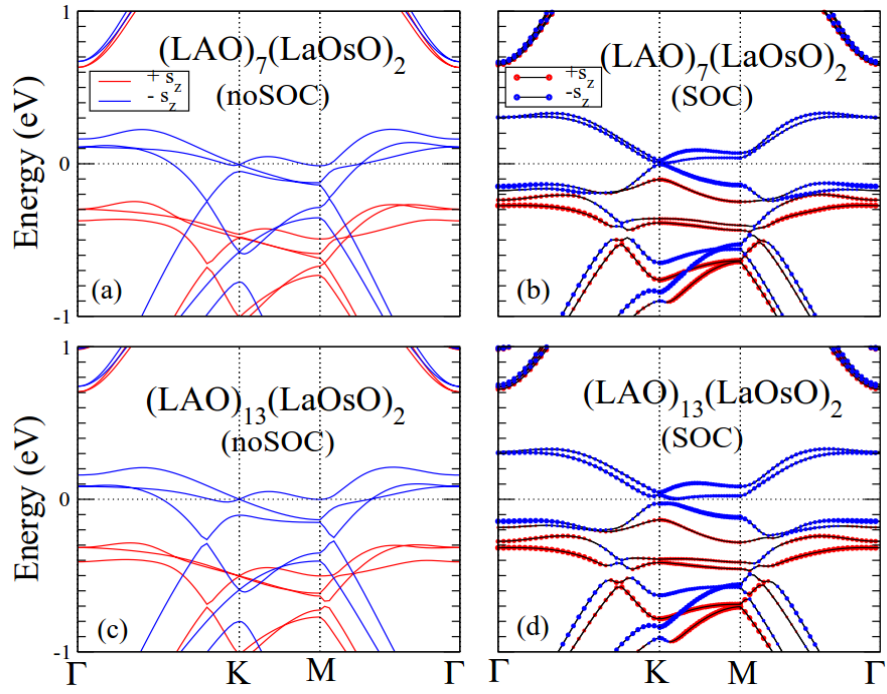

Figure S3. Band structures without (a,c) and with (b,d) SOC for the (LaOsO)$_2$/(LAO)$_7$ (a,b) and (LaOsO)$_2$/(LAO)$_{13}$ (c,d) superlattices. The Fermi level is at 0 eV.